\documentclass[aps, prb, twocolumn, noeprint]{revtex4-2}

\usepackage{cmap}

\usepackage[english]{babel}
\usepackage{amsmath}
\usepackage{bm}

\usepackage{graphicx}
\usepackage[caption=false]{subfig}

\usepackage[bookmarks=false]{hyperref}
\hypersetup{pdfstartview={FitH top}}
\hypersetup{
    colorlinks=true,
    linkcolor=blue,
    citecolor=blue,
    urlcolor=blue
}

\begin{document}

\title{Dipole Ordering of Water Molecules in Cordierite: Monte Carlo Simulations}

\author{Veniamin A. Abalmasov}
\email{abalmasov@iae.nsc.ru}

\affiliation{Institute of Automation and Electrometry SB RAS, 630090 Novosibirsk, Russia}

\date{\today}





\begin{abstract}

Electric dipoles of water molecules, enclosed singly in regularly spaced nanopores of a cordierite crystal, become ordered at low temperature due to their mutual interaction and show the frequency dependence of their dielectric susceptibility, typical for relaxor ferroelectrics, according to recent experimental data. The corresponding phase transition is accompanied by anomalies in thermodynamic quantities, such as heat capacity and dielectric susceptibility, which are calculated here using the Monte Carlo method, and their agreement the experimental data is discussed. Despite the increase in the correlation length, the partially filled dipole lattice at low temperatures, according to the calculations, does not have long-range order and corresponds to a dipole glass. This simulation gives a microscopical insight into the formation of polar nanoregions in relaxor ferroelectrics and the temperature dependence of their size.

\end{abstract}

\maketitle

Water is one of the key substances in nature. Studying the behavior of water molecules in various molecular environments is important for life sciences \cite{ball2017, hoelzel2020}. 
Water molecules, due to their small size and large electric dipole moment, are also excellent model objects for studying the collective behavior of electric dipoles at the nanoscale. This is important for understanding the ferroelectric and relaxor properties of technologically important materials, which are determined by the appearance of polar nanoregions, the nature, properties, and very existence of which are still hotly debated \cite{lines2001, samara2001, cowley2011, kleemann2012, albarakaty2015, filipic2016, wang2016}. In addition, we can expect the appearance in such systems of new states of matter with exotic and potentially useful properties, such as the quantum electric dipole liquid \cite{shen2016}, an analogue of the magnetic spin liquid, which has been actively studied in recent years~\cite{sachdev2008, balents2010, rau2019}. 

Water molecules located in the pores of minerals such as beryl and cordierite are separated from each other by the host material at a distance of 5-10~\!\AA, which is much larger than their size of about 1~\!\AA. In this case, their mutual electrostatic interaction can be considered in the dipole approximation, and dipole ordering could be expected at low temperatures. Indeed, recent experiments have revealed the incipient ferroelectricity of water molecules in beryl,  which manifested itself in an increase and then saturation of the dielectric susceptibility at low temperatures \cite{gorshunov2016}. The absence of a phase transition was explained by quantum fluctuations, which are significant in beryl due to the shallow potential that determines six possible dipole directions of the water molecule in the pore. At the same time, in cordierite, the dielectric susceptibility due to water molecules has a frequency-dependent maximum at low temperatures~\cite{belyanchikov2019a, belyanchikov2019b, belyanchikov2020}, which is typical of relaxor ferroelectrics. This, together with the anomalies in the specific heat~\cite{belyanchikov2019b, belyanchikov2020, paukov2007} and the frequency of the optical phonon~\cite{kolesnikov2019} at the same temperature, can be considered as evidence of the phase transition of water molecule dipoles of the order-disorder type. 

The water molecules in cordierite are similar in many aspects to the inclusion molecules in some clathrates that were thoroughly studied some time ago~\cite{sixou1976}, in particular methanol molecules in $\beta$-quinol~\cite{woll2001, rheinstaedter2005}. However, in the latter case, the order-disorder phase transition of inclusion molecule dipoles was accompanied by a structural phase transition of the entire crystal, which suggested the participation of additional to the dipole interactions~\cite{sixou1976}.

\begin{figure*}[t]
\centering
\includegraphics[trim=0cm 2cm 0cm 2.5cm, clip, width=1 \textwidth]{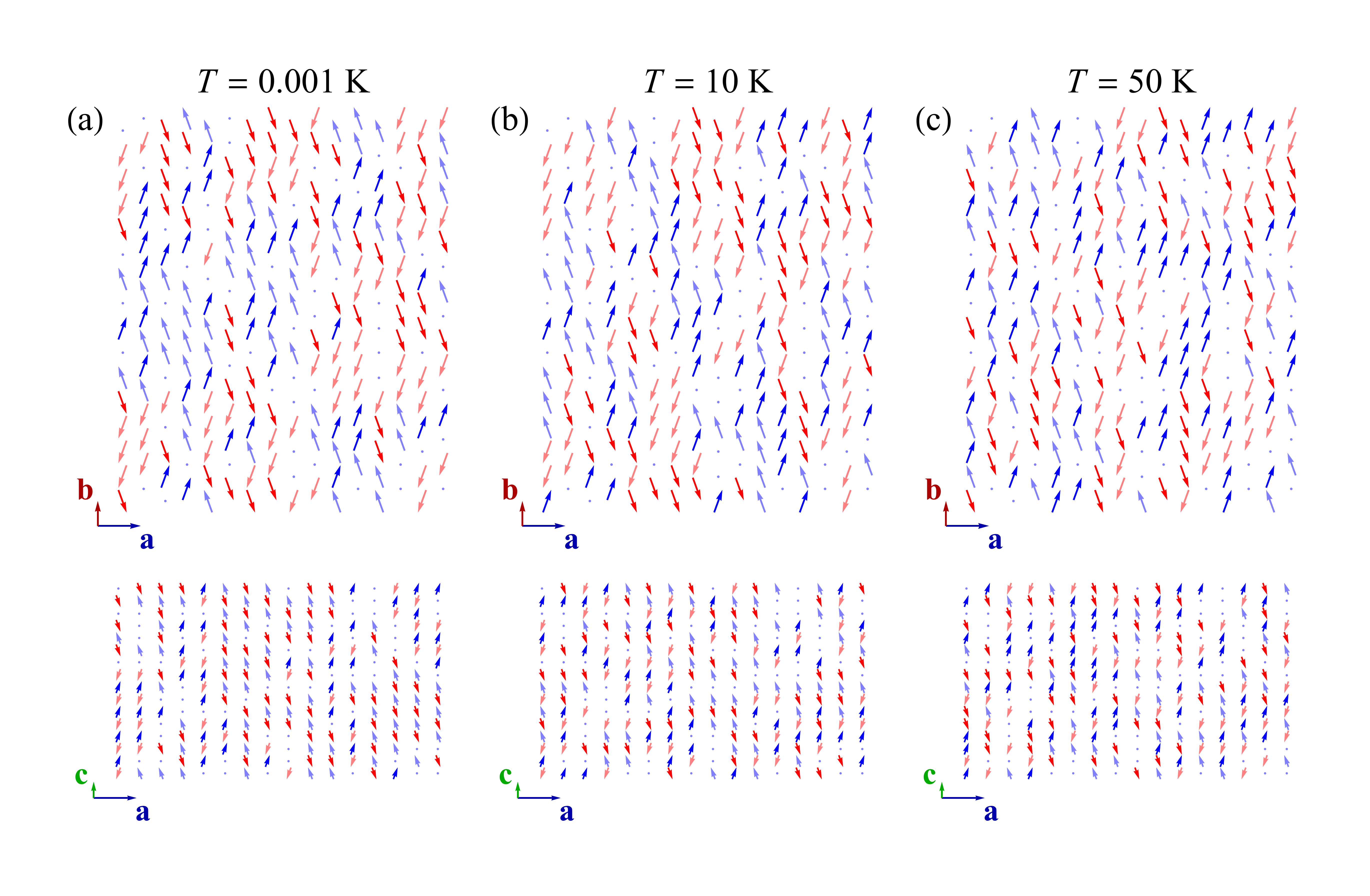}
\caption{Dipole configuration at the lowest, intermediate and highest temperature in calculations. Red and blue colors correspond to negative and positive dipole components along the $b$-axis, lighter and darker colors correspond to negative and positive dipole components along the $a$-axis. For convenience, the directions of the dipoles in the $ac$-panels are drawn so as to correspond to their directions in the $ab$-panels in the same place, since the dipoles have a zero component along the $c$-axis.}
\label{fig:dipoles}
\end{figure*}

In this letter I present the results of Monte Carlo simulations for electric dipoles of water molecules in cordierite, revealing the relationship between their thermodynamic properties and dipole configurations at different temperatures. Occupying the channel cavities of the cordierite crystal, (Mg,Fe)$_2$[Al$_4$Si$_5$O$_{18}$], water molecules form a partially filled stacked triangular lattice with sides of the isosceles triangle of $9.90$ and $9.74$~\AA\,  in the $ab$-plane and a distance of $4.66$~\AA\, between molecules along the $c$-axis \cite{kolesov2000, kolesnikov2014, belyanchikov2020} (Fig.~\ref{fig:dipoles}). In my calculations, I neglect the slight distortion of the triangle and assume that its side $b = 9.74$~\AA\, is exactly twice the distance $c$ between the neighboring dipoles along the $c$-axis. In the absence of alcali impurities, the H-H vector of water molecules is directed along the $c$-axis, and their dipole moments lie in the $ab$-plane, being dynamically disordered at high temperatures. However, the data on the directions of the dipole moment in the plane, which are determined by the interaction of the water molecule with the host crystal, are still controversial. The cavities in cordierite are anisotropic, smaller in size along the $b$-axis~\cite{belyanchikov2020}, and, according to different spectroscopic studies, the dipole of each water molecule can have two or four directions, including those related by inversion symmetry~\cite{kolesov2000}. Recent {\it ab-initio} calculations suggest four possible directions for a water molecule dipole in cordierite with an angle $\varphi$ between the dipole and $b$-axis of about~$10^{\circ}$~\cite{belyanchikov2020}. X-ray diffraction studies give a larger estimate for the angle of about $37^{\circ}$~\cite{dudka2020}. The four-directions hypothesis is also supported by the presence of the dielectric response of water molecules along both axes in the $ab$-plane \cite{belyanchikov2019a, belyanchikov2019b, belyanchikov2020}. The temperatures of about~3 and 15--30 K of the anomalies in the dielectric susceptibility along the $a$- and $b$-axes and the specific heat observed in the experiments \cite{belyanchikov2019a, belyanchikov2019b, belyanchikov2020, paukov2007} are reproduced in my Monte Carlo simulations with $\varphi = \pm 20^{\circ}$. Therefore, I use this angle in all the calculations reported here. So four possible dipole directions are given by the vector ${\bf p} = \pm p_0 (\sin \varphi, \cos \varphi, 0)$, where  $p_0$ is the absolute value of the dipole moment of the water molecule.

The energy of the dipole-dipole interaction of the water molecules is
\begin{align}\label{energy}
    E = \frac{k_e}{2} \sum_{n \neq m}^{N} \sum_{\alpha, \beta}  \frac{p^{\alpha}({\bf r}_n) \, p^{\beta}({\bf r}_m)}{|{\bf r}_{nm}|^3}\left(\!\delta^{\alpha \beta} - 3 \frac{r_{nm}^{\alpha}r_{nm}^{\beta}}{|{\bf r}_{nm}|^2}\!\right),
\end{align}
where $k_e = (4\pi \varepsilon_0 \varepsilon_r)^{-1}$, $\varepsilon_0$ is the electric constant, $\varepsilon_r$ is the relative permittivity of the cordierite matrix (which is considered isotropic here, as measured in \cite{belyanchikov2019a}, in contrast to \cite{ishai2020}, where it was assumed to be highly anisotropic). ${\bf r}_{nm} ={\bf r}_n - {\bf r}_m$ is the vector between two dipoles and ${\bf r}_n = {\bf R}_n + {\bf u}_n$ gives the position of the $n^{\text{th}}$ dipole, where ${\bf R}_n = n_1 {\hat {\bf a}} + n_2 {\hat {\bf b}} + n_3 {\hat {\bf c}}$ is the Bravais lattice vector, $n_{1,2,3}$ are integers, and ${\hat {\bf a}} = b \sqrt{3} {\hat {\bf x}}$, ${\hat {\bf b}} = b {\hat {\bf y}}$, ${\hat {\bf c}} = c \,{\hat {\bf z}}$ are primitive vectors. To obtain a rectangular sample, the Bravais lattice with two basis vectors, ${\bf u}_1 = (0, 0, 0)$ and ${\bf u}_2 = b (\sqrt{3}/2, 1/2, 0)$, is used (Fig. \ref{fig:dipoles}). The summation in Eq. (\ref{energy}) is over all dipole indices $n$ and $m$ in the sample, as well as the vector components $\alpha$ and $\beta$ along the coordinate axes.

In reality, not all pores are filled with water molecules, and empty pores represent site defects in the dipole lattice. In the absence of defects, the low temperature phase is expected to be antiferroelectric (AFE) with the propagation vector ${\bf k} = (0, 0, \pi/c)$, i.e. adjacent dipoles are antiparallel along the $c$-axis and parallel in the $ab$-plane. The coupling constant with the nearest neighbor along the $c$-axis is $J_{c1} = k_e p_0^2 / c^3 = 42.9$~K, expressed in temperature units, for $p_0 = 1.85$~D, $\varepsilon_r = 5$ \cite{belyanchikov2020}. The total in-plain coupling constant with six adjacent dipoles in the $ab$-plane is equal to $J_{ab}=(3/8) J_{c1}$. With the same accuracy in cordierite, the electric field created by the two nearest and two next-to-nearest dipoles along the $c$-axis with the above ${\bf k}$ is $J_{c}=(14/8) J_{c1}$. In the mean-field approximation (MFA), this leads to the temperature of the AFE phase transition $T_c = (17/8) J_{c1}$. To improve the approximation, given the small lattice constant ratio $c/b = 0.5$, the interaction along the $c$-axis can be treated exactly. This leads to an additional factor of $\log^{-1} (J_{c} / J_{ab}) = 0.65$ to the critical temperature, while the exact solution in 2D implies an even smaller correction factor of $\log^{-1} (2 J_{c} / J_{ab}) = 0.45$~\cite{scalapino1975}. Lattice anisotropy with domination of scalar interaction along the $c$-axis in Eq. (\ref{energy}) also results in quasi-independence of the two dipole components. Thus, the ordering of the two components $p^{a}$ and $p^b$ can be anticipated with a critical temperature of about $T_c \log^{-1} (2 J_{c} / J_{ab})$ multiplied by $\sin^2 \varphi = 0.12$ and $\cos^2 \varphi = 0.88$ respectively.

The influence of defects in MFA is manifested in the multiplication of $T_c$ by the filling factor $f$. However, it is known that a sufficient number of defects leads to the destruction of the ordered phase and the appearance of a disordered Imry-Ma phase at low temperatures \cite{berzin2021}. The formation of the glass state can be considered within the replica method \cite{kirkpatrick1978, xu1991}, which gives the critical temperature like MFA. The critical dipole concentration, below which the long-range order is impossible, is determined by the ratio square of the fluctuation field to the mean field and in the present case is about $f_{\text{cr}} = 0.62$. However, this value can be expected to be underestimated, since for a simple cubic lattice, the replica method gives $f_{\text{cr}} = 0.46$~\cite{xu1991}, while the Monte Carlo result is about~0.65~\cite{alonso2010}. The random local-field approximation~\cite{vugmeister1997} turns out to be in excellent agreement with the Monte Carlo result for the critical concentration in a simple cubic lattice~\cite{alonso2010}, but in the case of cordierite it does not predict an ordered state at all, as for a 2D square lattice with dipoles perpendicular to the plane, overestimating thermal fluctuations or due to a possible first-order phase transition in the latter cases with highly frustrating interactions. Thus, for the dipole concentration $f = 0.75$ as in the experiment \cite{belyanchikov2020}, I estimate the critical temperatures within MFA with the above corrections for the two dipole components as $T_c^a = 3.6$~K and $T_c^b = 27.2$~K.

\begin{figure}[t]
\center
\includegraphics[width=0.8 \columnwidth]{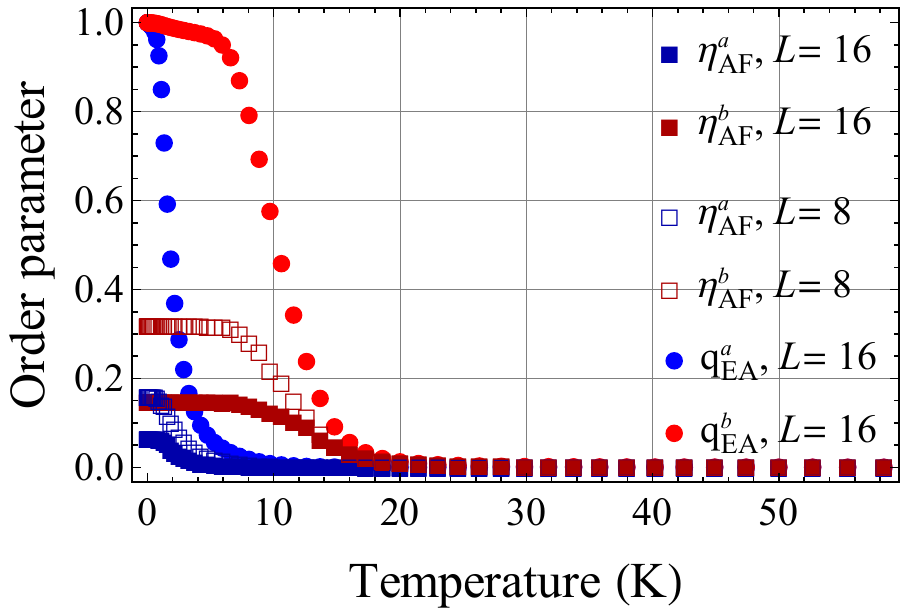}
\caption{Antiferroelectric, $\eta^{\alpha}_{\text{AF}}$, and Edwards-Anderson, $q^{\alpha}_{\text{EA}}$, order parameters for the sample side $L$ equal to 16 and 8.}
\label{fig:order}
\end{figure}

To study the behavior of the dipoles at different temperatures, single-spin-flip Monte Carlo simulations were performed using the standard Metropolis algorithm. Periodic boundary conditions were imposed, which in the case of the long-range dipole interaction imply the summation of an infinite array of image dipoles. This was done using the Ewald method, where summation was performed only in reciprocal space, and summation in real space was neglected due to the appropriate choice of the momentum-space cutoff parameter~\cite{wang2019}. The depolarizing term associated with surface charges was omitted, implying the usual experimental conditions with short-circuited boundaries of the entire macroscopic sample~\cite{wang2001}. The samples were rectangular with $L = 16$ lattice sites along each coordinate axis (4096 sites in total), unless otherwise indicated in specific cases. The filling factor of the dipole lattice $f = 0.75$, as in the experiment~\cite{belyanchikov2020}, yields $N = 3072$ dipoles in a sample. The results were averaged over 150 samples with different random defect configurations. $1,5 \times 10^4$ Monte Carlo steps per dipole (MCS) were used for collecting statistics with additional $10^3$ MCS for the equilibration at each temperature, descending from the highest temperature around 60~K to near zero about 0.001~K.

\begin{figure}[t]
\center
\includegraphics[width=0.8 \columnwidth]{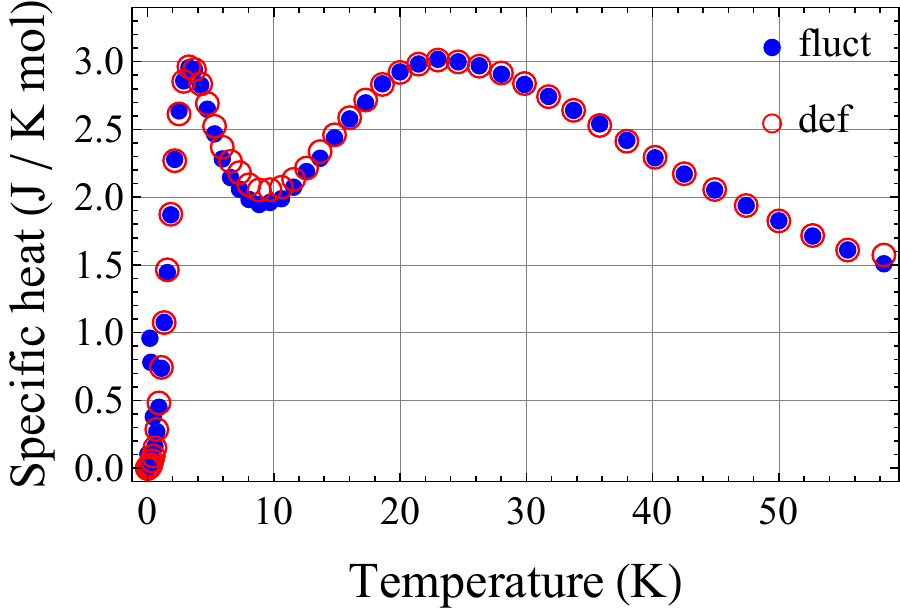}
\caption{Specific heat calculated from energy fluctuations (blue disks) and by definition (red circles).}
\label{fig:heat}
\end{figure}

In the absence of defects, the low temperature dipole order is AFE with the propagation vector ${\bf k} = (0, 0, \pi/c)$. The energy per dipole in this state is $E_{\text{AF}} = - 41.3$~K. Defects lead to the destruction of the long-range order (Fig. \ref{fig:dipoles}). At $f = 0.75$, the energy at zero temperature is $[E_{T=0}]_S = - 31.7$~K, while the energy of the AFE state is higher and equal to $[E_{\text{AF}}]_S = - 31.0$~K, where square brackets $[...]_S$ denote averaging over samples with different random defect configurations.

The AFE order parameter for each dipole component $p^\alpha$ is calculated as $\eta_{\text{AF}}^{\alpha} =(p^{\alpha})^{-1} [|\langle p^{\alpha}({\hat {\bf z}}\pi/c)\rangle_T|]_S$, where the dipole field Fourier transform is $p^\alpha ({\bf k}) = N^{-1} \sum_{n=1}^N p^\alpha ({\bf r}_n) \exp(i  {\bf k}\cdot{\bf r}_n)$, and angle brackets $\langle ... \rangle_T$ stand for thermal averaging. At first glance, it may seem that AFE phase transitions occur at approximately $T^a_{\text{AF}} = 4$~K and $T^b_{\text{AF}} = 16$~K, below which $\eta_{\text{AF}}^{\alpha}$ is not zero~(Fig. \ref{fig:order}). However, calculations for smaller samples with $L = 8$ (averaged over $10^5$ MCS and 250 samples) show an increase  in the saturation value of $\eta_{\text{AF}}^{\alpha}$ at low temperature, which evidences against the AFE phase transition \cite{alonso2010}. I note that the changes in other calculated thermodynamic quantities on going to $L = 8$ are not so significant. At the same time, the Edwards-Anderson glass order parameter, $q^{\alpha}_{\text{EA}}= (p^{\alpha})^{-2} [N^{-1}\sum_{n=1}^N \langle p_n^{\alpha}\rangle_T^2]_S$, shows that the relative number of frozen dipoles that do not flip during simulations at a given temperature increases rapidly below $T^a_{\text{EA}} = 6$~K and $T^b_{\text{EA}} = 16$~K for the corresponding dipole components (Fig. \ref{fig:order}), which may indicate a dipole-glass phase  transition at these temperatures \cite{vugmeister1990}.

The specific heat is calculated by definition as the derivative of energy with respect to temperature, $C = N_A N^{-1} d[\langle E \rangle_T]_S /dT$, and in terms of energy fluctuations as $C =N_ A (N k_B T^2)^{-1} [ \langle E^2 \rangle_T - \langle E \rangle_T^2 ]_S$, where $N_A$ is the Avogadro constant (Fig.~\ref{fig:heat}). It shows two maxima at about $T^a_{\text{cap}}=3$~K and $T^b_{\text{cap}} = 23$~K, and their smoothness also indicates the spin-glass transition \cite{alonso2010}. At several temperatures below 1~K, the specific heat values calculated from fluctuations are unreasonably high due to large rare energy fluctuations during the finite simulation time and can be considered as artifacts. At these points, the specific heat calculated by definition can be negative.

\begin{figure}[t]
\center
\includegraphics[width=0.8 \columnwidth]{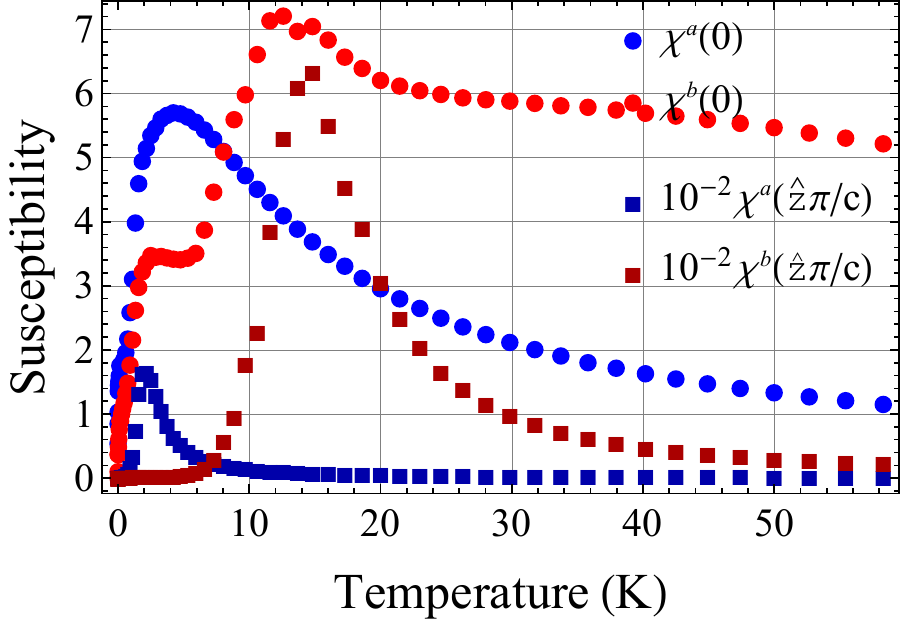}
\caption{Homogeneous, $\chi^{\alpha}(0)$, lighter colors and staggered (multiplied by $10^{-2}$), $\chi^{\alpha} ({\hat {\bf z}}\pi/c)$, in darker colors dielectric susceptibilities due to the dipole component $p^{\alpha}$.}
\label{fig:suscept}
\end{figure}

The polarization $P^{\alpha}({\bf k})$ and susceptibility $\chi^{\alpha}({\bf k})$ in the $ab$-plane ($\alpha = a, b$), both homogeneous (${\bf k} = 0$) and staggered along the $c$-axis (${\bf k} = (0, 0, \pi/c)$) are calculated as $P^{\alpha}({\bf k}) = v^{-1} p^{\alpha}({\bf k})$, where $v$ is the unit cell volume, and $\chi^{\alpha}({\bf k}) = N (k_B T)^{-1}[ \langle {P^{\alpha}({\bf k})}^2 \rangle_T - \langle {P^{\alpha}({\bf k})} \rangle_T^2 ]_S$. The staggered susceptibility $\chi^{a,b} ({\hat {\bf z}}\pi/c)$ has an anomaly at about $T^a_{\chi(c)} = 2$~K and $T^b_{\chi(c)} = 14$~K, while for the homogeneous susceptibility $\chi^{a,b} (0)$, $T^a_{\chi(0)} = 4$~K and $T^b_{\chi(0)} = 2, 12, 15$~K~(Fig.~\ref{fig:suscept}). In the absence of defects, $\chi^{b} (0)$ has only a very smooth maximum and decreases rapidly below the maximum temperature of $\chi^{b} ({\hat {\bf z}}\pi/c)$, which is typical for the antiferroelectric phase transition. This means that the sharp peaks of $\chi^{b} (0)$ arise solely from defects. At high temperatures, both susceptibilities follow the Curie-Weiss temperature dependence with the Curie-Weiss temperature of about $T^a_{\text{CW}} \approx \pm 5$~K and $T^b_{\text{CW}} \approx \pm 30$~K, positive for the staggered and negative for the homogeneous susceptibility.

The space correlation function for each dipole component $p^\alpha$ along the direction given by  the Bravais lattice vector ${\bf r}_n$ is calculated as
\begin{align}\label{correlation}
    C^{\alpha}&({\bf r}_n) = (1 -  (\eta_{\text{AF}}^{\alpha})^{2})^{-1} 
    \left([|\langle (f N)^{-1} (p^{\alpha})^{-2} \right.\nonumber \\
    & \times \sum_{m=1}^N \left. p^{\alpha}({\bf R}_m) p^{\alpha}({\bf R}_m + {\bf r}_n)\rangle_T|]_S - (\eta_{\text{AF}}^{\alpha})^2   \right),
\end{align}
when ${\bf r}_n \neq 0$ and  $C^{\alpha}(0) = 1$. The filling factor $f$ is introduced in Eq. (\ref{correlation}) to make the result independent of the dipole concentration.  The correlation along the $a$-axis is hardly visible over the statistical noise. For the other axes, it turns out to be best fitted by a simple exponent, which corresponds to the Ornstein-Zernike form for the asymptotic behavior of the correlation function \cite{landau2009, ott2018, pelissetto2002}, $C(r) \propto r^{-(d-1)/2} \exp(-r/\xi)$, in one dimension. The rather unexpected value of $d = 1$, for the three-dimensional dipole lattice, is probably due to the anisotropic nature of the dipole interaction (the correlation length is much larger in the direction of the dipole component) and the lattice itself (the correlation grows faster along the c-axis, along which the distance between the dipoles is minimal).

\begin{figure}[t]
\center
\includegraphics[width=0.8 \columnwidth]{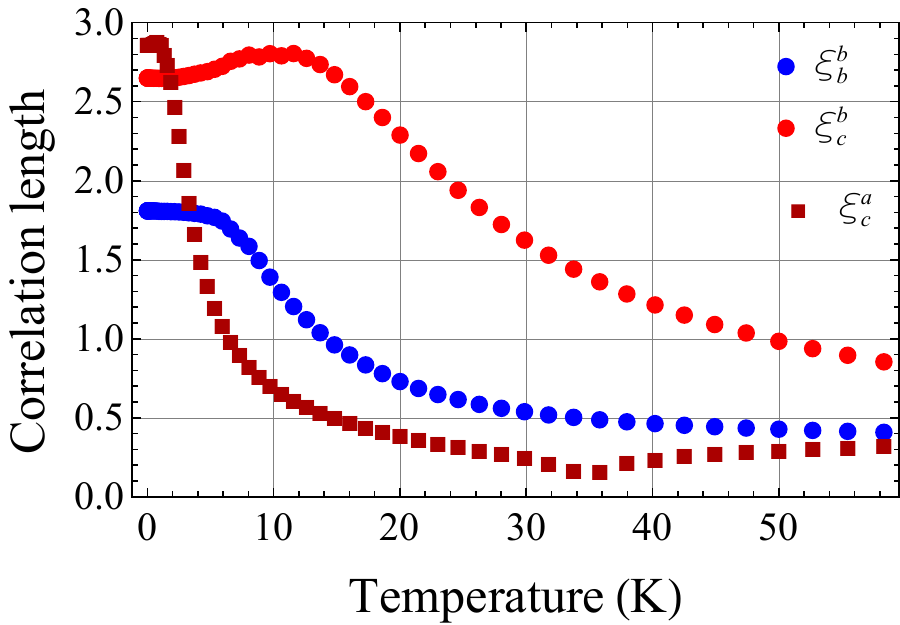}
\caption{Correlation length, $\xi^{\alpha}_{\beta}$, of the dipole vector component $p^{\alpha}$ along the $\beta$-axis in units of the dipole lattice constant along the $\beta$-axis.}
\label{fig:correlation}
\end{figure}

The corresponding correlation length, $\xi^{\alpha}_{\beta}$, where the superscript denotes the dipole vector component, and the lower one is the axis along which the length is measured, is calculated as $\xi^{\alpha}_{\beta} = - r^{\beta}_{n} / \ln C^{\alpha}(r^{\beta}_{n})$  \cite{pelissetto2002}. For the correlation along the $b$-axis, it is sufficient to take $n = 1$, for larger $n$ it is more prone to statistical errors. The correlation length along the $c$-axis, however, is better to calculate as $\xi^{\alpha}_{\beta} = 1 / \ln (C^{\alpha}(r^{\beta}_{n}) / C^{\alpha}(r^{\beta}_{n+1}))$, where the correlation function is taken at two successive sites of the Bravais lattice along the $\beta$-axis (with $n = 1$ as well), for the following reason. At low temperatures, the dipoles inside a cluster bounded by defects on both sides along the $c$-axis are completely ordered (Fig. \ref{fig:dipoles}(a)), which implies $C^{\alpha}(1\cdot {\hat {\bf c}}) = 1$ with its logarithm being zero. The temperature dependence of the thus obtained correlation lengths is shown in Fig.~\ref{fig:correlation}. Calculations with other reasonable values of $n$ or fitting the correlation function do not significantly change this result.

The correlation lengths are inversely proportional to the temperature above the phase transition temperature, saturate around it, and then have a finite value below it, which corresponds to the frozen short-range dipole order (Fig.~\ref{fig:correlation}). At high temperatures, they have a small nonzero value due to statistical errors in calculations together with a nonzero value of the lattice constant. A slight decrease in $\xi^{b}_{c}$ below the critical temperature can be associated with the onset of the AFE order parameter $\eta^{b}_{\text{AF}}$, see Eq. (\ref{correlation}). The slight kink in $\xi^{a}_{c}$ between 30 and 40~K is probably due to the presence of defects. At low temperatures, the correlation lengths  $\xi^{\alpha}_{c}$ along the $c$-axis are determined by the defect concentration and are about half the average size of the cluster, which is bounded by defects at both ends along this axis. At the same time, the correlation length $\xi^{b}_{b}$ is slightly more than half the average cluster size determined by a continuous sequence of the same value of the dipole component $p^b$ along the $b$-axis calculated in \cite{belyanchikov2020} using the Monte Carlo method with free boundary conditions.

The Monte Carlo simulation results for periodic boundary conditions presented here are very close to those obtained earlier for free boundary conditions~\cite{belyanchikov2020}. Meanwhile, the phase transition temperatures that appear in the Monte Carlo simulation turns out to be about four times lower than those predicted by simple MFA, and about twice as low when the interaction along the c axis is considered exactly as estimated above. This is not surprising, however, since MFA is known to significantly overestimate the critical temperature, especially in low dimensions or for competing interactions~\cite{abalmassov2019}.  At the same time, the replica method based on MFA underestimates the critical dipole concentration for the cordierite lattice, as was the case for the simple cubic lattice mentioned above, since it was proved here that for $f = 0.75$ the AFE order parameter vanishes for large sample sizes~(Fig. \ref{fig:order}).

The obtained results of the Monte Carlo simulation are consistent with the available experimental data.  Indeed, one can observe a wide anomaly in the heat capacity at a temperature of about 30~K in the experimental data~\cite{belyanchikov2020, paukov2007}, which corresponds well to the broad peak in Fig.~\ref{fig:heat} due to the ordering of the $p^b$-component of the dipoles. At the same time, the low temperature peak at about 3~K in Fig.~\ref{fig:heat} is strongly flattened in the experimental data~\cite{belyanchikov2020, paukov2007}, although it is quite visible in $C/T$ data in \cite{belyanchikov2019b}. In principle, this could be due to the possible tunneling of water molecules between two states with the same $p^b$-component, which are separated by a smaller angle $2\varphi$ (and, most likely, by a smaller energy barrier). The maximum of the dielectric susceptibility $\chi^a(0)$ at a temperature of about 3~K in Fig.~\ref{fig:suscept} is in good agreement with the peak of the dielectric susceptibility obtained by fitting the soft mode~\cite{belyanchikov2020} and with the kink of the dielectric measurement data in~\cite{belyanchikov2019a, belyanchikov2019b, belyanchikov2020}. The position of the broad maximum of $\chi^b(0)$ at about 30~K in~\cite{belyanchikov2019a, belyanchikov2019b} is also close to that in Fig.~\ref{fig:suscept}. Although the experimental values of $\chi^b(0)$ appear to be about half the calculated ones. Finally, an increase in the correlation length (Fig.~\ref{fig:correlation}) can explain the frequency dependence of the dielectric susceptibility observed in \cite{belyanchikov2020, belyanchikov2019a, belyanchikov2019b}, as in the case of one-dimensional single-chain magnets \cite{zhang2013}.

The singularities in the staggered dielectric susceptibilities $\chi^{\alpha} ({\hat {\bf z}}\pi/c)$ (Fig.~\ref{fig:suscept}) imply softening of the polar relaxation mode at the boundary of the Brillouin zone, which can be measured experimentally by inelastic neutron ans x-ray scattering techniques~\cite{milesibrault2020}. The temperature dependence of the correlation lengths of the dipole components (Fig.~\ref{fig:correlation}), in turn, can be measured in x-ray and neutron diffraction experiments~\cite{rheinstaedter2005}. Together, this could be a crucial test for the model of water molecules in cordierite considered here. 

In conclusion, the Monte Carlo simulations performed within the four-directions model for dipoles of water molecules in cordierite revealed anomalies in heat capacity and dielectric susceptibility in accordance with the available experimental data \cite{belyanchikov2019a, belyanchikov2019b, belyanchikov2020, paukov2007}. It was shown that site defects in the dipole lattice of water molecules destroy the long-range dipole order at low temperatures. At the same time, with decreasing temperature, the short-range order gradually appears, which can reflect the formation of the polar nanoregions in relaxors, since the correspondence between relaxors and dipole glasses is generally  recognized~\cite{lines2001, samara2001, cowley2011, kleemann2012, albarakaty2015, filipic2016, wang2016}. 

I thank B. E. Vugmeister for useful discussions.

The reported study was funded by RFBR, project number 20-02-00314.


%

\pagebreak

\end{document}